\documentclass[aps,twocolumn,pra]{revtex4}
\usepackage[ansinew]{inputenc}
\usepackage{amsmath}
\usepackage{graphicx}

\begin{document}
%\twocolumn[
\title{Preparation and probing of the ground state coherence in Rubidium  }
\author{Martin Oberst\cite{oberst}}
\address{Fachbereich Physik, Universit\"at Kaiserslautern, 67653
Kaiserslautern, Germany}
\author{Frank Vewinger\cite{vewinger} and A. I. Lvovsky}
\address{Institute for Quantum Information Science, University of
Calgary, Alberta T2N 1N4, Canada}
\begin{abstract}
We demonstrate the preparation and probing of the coherence between
the hyperfine ground states \mbox{$|5S_{1/2}, F=1\rangle$} and
$|5S_{1/2}, F=2\rangle$ of the Rubidium 87 isotope. The effect of
various coherence control techniques, i.e. fractional Stimulated
Raman Adiabatic Passage and Coherent Population Return on the
coherence are investigated. These techniques are implemented using
nearly degenerate pump and Stokes lasers at 795nm (Rubidium D1
transition) which couple the two hyperfine ground states via the
excited state $|5P_{1/2}, F=1\rangle$ through a resonant two-photon
process, in which a coherent superposition of the two hyperfine
ground states is established. The medium is probed by an additional
weak laser, which generates a four-wave mixing signal proportional
to the ground state coherence, and allows us to monitor its
evolution in time. The experimental data are compared with numerical
simulations.
\end{abstract}
\maketitle
%\ocis{190.4380,190.2620,190.1900,270.1670,300.6210}
%]

\textbf{Introduction} The techniques of Stimulated Raman Adiabatic
Passage (STIRAP) \cite{StirapRev} and Coherent Population Return
(CPR) \cite{Har2} are of importance in various fields of physics.
STIRAP is used for, e.g., coherent control of population transfer
\cite{StirapRev} between atomic levels, preparing coherent
superposition states \cite{Vew1}, efficient generation of teraherz
radiation\cite{Kal} and probing atom-photon entanglement\cite{Vol1}.
Coherent population transfer helps to enhance frequency conversion
processes\cite{Man1}, generate attosecond
pulses\cite{Har31,Har32,Har33,Har34,Har35,Har36}, and improve the
resolution of spectroscopic measurements\cite{Half1}. Most
investigations of these techniques have so far concentrated on the
dynamics of \emph{population} of various energy levels in the system
but never on the \emph{coherence} between them. At the same time,
coherence is an essential element of these processes, and it is
important to develop a technique for its time-resolved measurement.
The present paper achieves this goal.

In this work we prepare and probe the coherence between the two
hyperfine ground states $|5S_{1/2}, F=1\rangle$ and $|5S_{1/2},
F=2\rangle$ in the $^{87}$Rb isotope. We perform, for the first
time, a detailed study of two coherence manipulation techniques,
namely fractional STIRAP\cite{Vit2} and CPR. For both these
processes, the temporal evolution of the coherence is investigated
and compared to numerical simulations.

\begin{figure}[htb]
\centerline{\includegraphics[width=7cm,height=3.5cm]{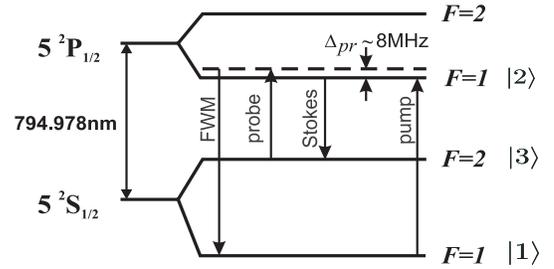}}
  \caption{\small Coupling scheme for the preparation and probing of the coherence $\rho_{13}$ in Rubidium 87.
  The hyperfine splittings and detunings are exaggerated for clarity.}
  \label{figure1}
\end{figure}

 \textbf{Theory} Both schemes use two laser fields, referred to as
the pump and Stokes fields, which resonantly couple the hyperfine
state $|1\rangle$ (5$S_{1/2}$, $F=1$) of $^{87}$Rb to an
intermediate state $|2\rangle$ (5$P_{1/2}$, $F=1$) and state
$|2\rangle$ to the target state $|3\rangle$ (5$S_{1/2}$, $F=2$) in a
$\Lambda$-type energy level configuration (Fig.~\ref{figure1}).
Neglecting decoherence, the coupling creates a (transient or
permanent) coherent superposition between states $|1\rangle$ and
$|3\rangle$, the so-called dark state\cite{StirapRev} of the system:
\begin{equation}
|ds\rangle=\cos\theta(t) |1\rangle - \sin\theta(t) |3\rangle,
\label{eins}
\end{equation}
where the mixing angle $\theta(t)$ is determined by
\begin{equation}\label{tantheta}
\tan\theta(t)=\Omega_P(t)/\Omega_S(t)
\end{equation}
 with $\Omega_{P,S}$ being the Rabi frequencies
of the pump and Stokes lasers.

The coherence between states $|1\rangle$ and $|3\rangle$ is defined
as the off-diagonal element $\rho_{13}$ of the system's density
matrix. According to Eq.~(\ref{eins}), it is given by:
\begin{equation}
\rho_{13}(t)=\cos\theta(t)\,\sin\theta(t). \label{zwei}
\end{equation}
and thus is fully determined by the temporal evolution of the Rabi
frequencies $\Omega_P$ and $\Omega_S$. For fractional STIRAP, the
pump and Stokes lasers are pulsed, exciting the medium in a
counter-intuitive order, i.e. the Stokes laser precedes the pump. At
the moment  $t=t_f$ both lasers are simultaneously, adiabatically
turned off, resulting in a constant ratio (\ref{tantheta}) for $t >
t_f$. Any coherence $\rho_{13}$ prepared prior to $t = t_f$ will
then stay in the system even after the termination of the driving
fields\cite{Vit2}, gradually dephasing at the ground state
decoherence rate\cite{Fige}.

The second technique used to prepare coherence $\rho_{13}$ is
coherent population return. Here the pump field is continuous
($\Omega_P={\rm const}$) and the Stokes laser is pulsed with a
Gaussian shape, so $\Omega_S(t)$ is time dependent. The atomic
population, initially in state $|3\rangle$, is transferred to state
$|1\rangle$ and back again as the Stokes field is applied, resulting
in a transient coherence between the two states.

The ground state coherence is probed by an additional weak laser
field, which induces a four-wave mixing (FWM) signal close to the
pump laser wavelength at 795nm (Fig.~1). Assuming perfect phase
matching and neglecting depletion, absorption and losses, which is
justified for the optical densities used, the propagation of the FWM
field is given by:
\begin{eqnarray}
\left|\frac{\partial}{\partial z}E_{\text{FWM}}\right|=\frac{\pi N
\omega_{\text{FWM}}\mu_{13}\mu_{23}}{\hbar c \epsilon_{0}
\Delta_{pr}}|\rho_{13}||E_{pr}|.\label{coherence1}
\end{eqnarray}
Here $N$ denotes the number density of the sample, $\mu_{23}$ and
$\mu_{13}$ are the dipole moments for the corresponding transitions,
$E_{pr}$ is the probe field amplitude, $\Delta_{pr}$ is the detuning
of the probe field from the transition
$|2\rangle\rightarrow|3\rangle$. All fields are assumed to propagate
in the positive $z$ direction. Equation (\ref{coherence1}) shows
that the electric field of the FWM radiation at the frequency
$\omega_{FWM}$ is proportional to the coherence $|\rho_{13}|$ and
can serve to measure the latter.

 \textbf{Experimental setup} The experiments are carried out in a
Rubidium vapor cell at a temperature of 60$°$C, which corresponds to
a number density on the order of $10^{11}$ cm$^{-3}$. The pump field
is generated by a continuous Ti:Sapphire laser (Coherent MBR-110)
with a narrow spectral line width ($\sim$100 kHz). The Stokes and
probe fields are both provided by a continuous diode laser (Toptica
DLL100) which is frequency locked to the Ti:Sapphire laser at the
6.8-GHz Rubidium hyperfine splitting of the $|5^2S_{1/2}\rangle$
state with a precision of better than 10 Hz. The output of the diode
laser is split into the Stokes and probe beams via a combination of
a $\lambda/2$ wave plate and a polarizing beam splitter.

All the three beams are sent through acousto-optical modulators,
which allows control over the temporal shape, frequency and relative
position of the different pulses in time. We employ the single-pass
configuration for the Stokes and probe fields, resulting in a
frequency shift of 80~MHz and 88~MHz, respectively, and the
double-pass configuration for the pump leading to a 160~MHz shift.
The three beams with diameters (full-width at half-maximum, FWHM) of
\mbox{$\o_P$=1.26 mm}, $\o_S$=1.76 mm and $\o_{pr}$=0.76 mm for
pump, Stokes and probe, respectively,  are then overlapped and
directed into the Rb vapor cell located in a magnetically shielded
oven. The cell contains 13 millibars of Neon as a buffer gas. The
polarizations of the probe laser and the generated FWM signal are
orthogonal to the pump and Stokes lasers.

After the cell, the FWM and probe are separated from the other
fields using a polarizer and overlapped with the unmodulated output
of the Ti:Sapphire laser on a fast photodiode for heterodyne
detection. The beat signal proportional to the electric field of the
FWM signal is monitored as a function of time using a spectrum
analyzer (Agilent ESA 9905) with a 200-ns resolution.

\begin{figure}[htb]
\centerline{\includegraphics[width=7.5cm,height=5cm]{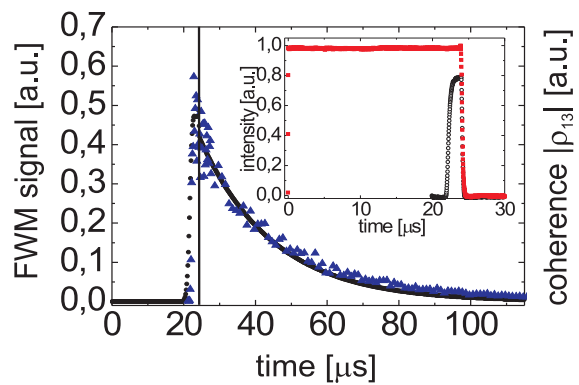}}
 \caption{\small Fractional STIRAP process: FWM signal as a function of the probe laser delay
 $\Delta\tau_{pr}$ (triangles) and numerical simulation (solid
 circles). Inset: temporal shape of the pump (hollow circles) and Stokes
 pulses (solid rectangles).} \label{figure2}
\end{figure}

 \textbf{Results and discussion} In order for fractional STIRAP to work, the population has
to be transferred to state $|1\rangle$ at the beginning of the
process. This is realized by optical pumping with the leading part
of the Stokes pulse which exceeds the duration of the pump pulse by
more than one order of magnitude ($\tau_{S}\approx24$ $\mu$s and
$\tau_{P}\approx 2$ $\mu$s, respectively (FWHM of intensity)); see
Fig.~\ref{figure2}. The probe laser is pulsed with a pulse duration
of $\tau_{pr}$=500 ns (FWHM of intensity). Typical field powers are
on the order of several mW for the pump and Stokes radiation and a
few hundred $\mu$W for the probe laser. We verified that the probe
laser does not significantly influence the coherence by checking
that the FWM intensity is proportional to that of the probe. With
the pulse parameters given, we estimate the Rabi frequencies to be
on the order of $\Omega_P=0.09$ ns$^{-1}$, $\Omega_S=0.12$ ns$^{-1}$
and $\Omega_{pr}=0.08$ ns$^{-1}$ respectively for the pump, Stokes
and probe lasers.

To measure the coherence $\rho_{13}$, the delay $\Delta\tau_{pr}$ of
the probe pulse with respect to the pump and Stokes pulses is varied
in 0.2 $\mu$s steps and the FWM signal is detected.
Fig.~\ref{figure2} shows the time dependence of the FWM radiation
and therefore the temporal evolution of the prepared coherence
$|\rho_{13}|$ in the system.

The presence of coherence and FWM after the excitation lasers have
been turned off contrasts with the classically expected behavior of
four-wave mixing, where the created field is proportional to the
product of the three fields irradiated on the medium,
$E_{\text{FWM}}\propto E_SE_PE_{\text{pr}}$. This residual coherence
$\rho_{13}$ is a signature of fractional STIRAP \cite{Vit2} since
only a part of the population is transferred from state $|1\rangle$
to state $|3\rangle$ during the interaction with pump and Stokes
laser pulse. Upon termination of the excitation pulses, the signal
experiences exponential decay due to decoherence \cite{Fige}. We
found the decay constant to equal 20 $\mu$s.

For the CPR experiment, the system is prepared in state $|3\rangle$
by means of optical pumping via a continuous pump laser with a Rabi
frequency of $\Omega_P \approx 0.028$ ns$^{-1}$ (as determined from
the beam parameters). The Stokes laser is pulsed with a Gaussian
pulse shape of a \mbox{$\tau_{S}$=15.3 $\mu$s} duration (FWHM of
Rabi frequency). We perform several experimental runs with different
Stokes field intensities. The probe laser is provided as continuous
radiation with $\Omega_{\text{pr}}\approx 0.006\ \text{ns}^{-1}$(as
determined from the beam parameters). The continuous probe laser
provides real-time information on the evolution of coherence. As in
the previous case, the probe laser does not significantly influence
the prepared coherence.

\begin{figure}[htb]
\centerline{\includegraphics[width=7.5cm,height=8.5cm]{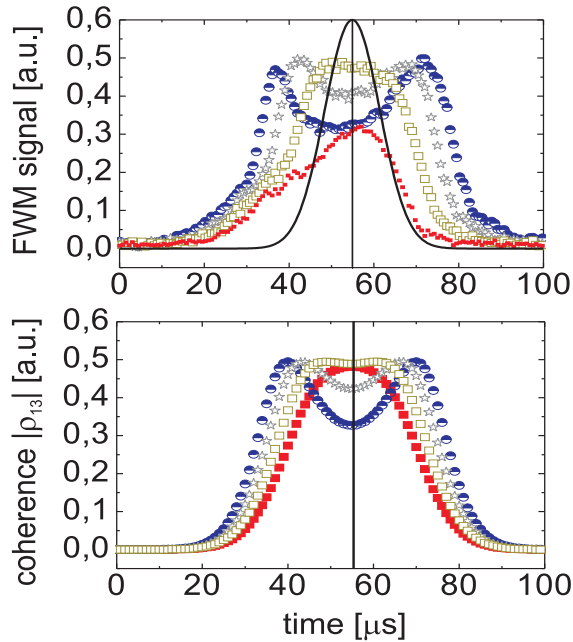}}
\caption{\small Coherence during the CPR process: a) The measured
FWM signal and b) calculated coherence $|\rho_{13}|$ for different
Stokes pulse intensities. The maximum Stokes Rabi frequencies (at
$t=54.9\ \mu$s) are: $\Omega_{S}$=0.023 ns$^{-1}$ (solid
rectangles); $\Omega_{S}$=0.033 ns$^{-1}$ (hollow rectangles);
$\Omega_{S}$=0.05 ns$^{-1}$ (stars); $\Omega_{S}$=0.075 ns$^{-1}$
(half solid circles). The solid line in (a) shows the Stokes pulse
intensity profile (scale given in arbitrary units).} \label{figure3}
\end{figure}

Figure \ref{figure3}(a) shows the measured FWM signal as a function
of the Stokes Rabi frequency. This result is in good agreement with
the numerically calculated time evolution of the coherence
$|\rho_{13}|$ [Fig. \ref{figure3}(b)]. The only fit parameter used
in the calculation is the proportionality coefficient between the
Stokes Rabi frequency and the square root of the laser intensity.
The discrepancy present in the raising edges of the curves is due to
absorption of the Stokes laser in the vapor cell. The same reason
brings about the asymmetry in the measured shapes.

Again, the signal observed in the CPR experiment shows significantly
nonclassical behavior. Whereas a classical FWM should be
proportional to the Stokes field, the signal measured at
$\Omega_S>\Omega_P$ exhibits a minimum when the Stokes field is
maximized. This is easily understood: according to Eq.~(\ref{zwei}),
the maximum coherence is reached when $\Omega_S(t)=\Omega_P$  and
both levels $|1\rangle$ and $|3\rangle$ are equally populated. If
the Stokes Rabi frequency increases above that of the pump, and thus
$\theta$ falls below 45$^\circ$, the coherence is reduced, as more
than a half of the population is transferred to state $|1\rangle$.
This behavior is clearly displayed by the experimental observation,
Fig.~\ref{figure3}.

 \textbf{Conclusion} We have demonstrated the preparation of the
ground state coherence in Rubidium vapor via fractional STIRAP and
CPR, and the detection of this coherence by four mave mixing. The
experimental results fit very well with numerical simulations and
give clear evidence for the nonclassical character of these
coherence control techniques. The presented method provides a
powerful tool for the preparation and probing of coherence in a
variety of systems.

\textbf{Acknowledgement} We acknowledge valuable discussions with K.
Bergmann, T. Halfmann and E. Figueroa and J. Appel for assistance
with the laser system. The work was funded by the Deutsche
Forschungsgemeinschaft (Graduiertenkolleg 792), AIF, CFI, CIAR,
NSERC, and Quantum$Works$.


\begin{thebibliography}{99}
\bibitem[*]{oberst} Email: oberst@physik.uni-kl.de
\bibitem[\dag]{vewinger} Present address: Institut f\"ur angewandte Physik, Universit\"at
Bonn, Wegelerstr. 8, 53115 Bonn, Germany.
\bibitem{StirapRev} N. V. Vitanov, T. Halfmann, B. W. Shore, and K. Bergmann, {Annu. Rev. Phys. Chem.} {\bf 52} 764 (2001).
\bibitem{Har2} M. Jain and S. E. Harris, {\em Opt. Lett.} {\bf 22} 636 (1997);
\bibitem{Mer2} A. J. Merriam, S.J. Sharpe, H. Xia, D. Manuszak, G. Y. Yin, and S. E. Harris, {\em Opt. Lett.} {\bf 24} 625 (2001).
\bibitem{Vew1} F. Vewinger, M. Heinz, R. Garcia-Fernandez, N.V. Vitanov, and K.
Bergmann, {\em Phys. Rev. Lett.} {\bf 91} 213001 (2003).
\bibitem{Kal} N. G. Kalugin, and Y. V. Rostovtsev, {\em Opt. Lett.} {\bf 31} 969 (2006).
\bibitem{Vol1} J. Volz, M. Weber, D. Schlenk, W. Rosenfeld, J. Vrana, K. Sauke, C. Kurtsiefer and H.
Weinfurter, {\em Phys. Rev. Lett.} {\bf 96} 030404 (2006).
\bibitem{Man1} M. Jain, H. Xia, G.Y. Yin, A. J. Merriam, and S. E. Harris, {\em Phys. Rev. Lett.} {\bf 77} 4326 (1996).
\bibitem{Har31} S. E. Harris and A. V. Sokolov, Phys. Rev. A {\bf 55}, 4019 (1997)
\bibitem{Har32} A. V. Sokolov, D. R. Walker, D. D. Yavuz, G. Y. Yin, and S. E. Harris, {\em Phys. Rev. Lett.}
{\bf 85}, 562 (2000);
\bibitem{Har33} A. V. Sokolov, D. D. Yavuz, D. R. Walker, G. Y. Yin, and S. E. Harris, {\em Phys. Rev. A} {\bf
63}, 051801 (2001);
\bibitem{Har34} S. E. Harris and A. V. Sokolov, {\em Phys. Rev. Lett.} {\bf 81}, 2894 (1998);
\bibitem{Har35} A. V. Sokolov, D. D. Yavuz, and  S. E. Harris, {\em Opt. Lett.} {\bf 24}, 557 (1999);
\bibitem{Har36} D. D. Yavuz, A. V. Sokolov, and
S. E. Harris, {\em Phys. Rev. Lett.} {\bf 84}, 75 (2000);
\bibitem{Half1} T. Halfmann, T. Rickes, N. V. Vitanov, and K. Bergmann, {\em Opt. Commun.} {\bf 220} 353 (2003).
\bibitem{Vit2} N. V. Vitanov, K. A. Suominen and B. W. Shore, {\em J. Phys. B} {\bf 32} 4535 (1999)
\bibitem{Fige} E. Figueroa, F. Vewinger, J. Appel and A. I. Lvovsky, Opt. Lett. {\bf 31}, 2625 (2006)
\end{thebibliography}
\end{document}